\documentclass[prl,floatfix,showpacs,twocolumn,superscriptaddress]{revtex4}
\usepackage{amsfonts}
\usepackage{amssymb}
\usepackage{graphicx}
\begin{document}

\title{Unified Description of Aging and Rate Effects in Yield of Glassy Solids}
\author{J\"{o}rg Rottler} 
\affiliation{Princeton Institute for the Science and Technology
of Materials (PRISM), Princeton University, Princeton, New Jersey
08544}
\author{Mark O.~Robbins} 
\affiliation{Dept. of Physics and Astronomy, Johns Hopkins University,
 3400 N. Charles St., Baltimore, MD 21218}

\date{\today}

%\begin{document}
%\maketitle

\begin{abstract} 
The competing effects of slow structural relaxations (aging) and
deformation at constant strain rate on the shear yield stress $\tau^y$
of simple model glasses are examined using molecular simulations.  At
long times, aging leads to a logarithmic increase in density and
$\tau^y$.  The yield stress also rises logarithmically with rate, but
shows a sharp transition in slope at a rate that decreases with
increasing age.  We present a simple phenomenological model that
includes both intrinsic rate dependence and the change in properties
with the total age of the system at yield.  As predicted by the model,
all data for each temperature collapse onto a universal curve.
\end{abstract}

\maketitle 

The mechanical behavior of amorphous materials such as polymers
\cite{haward1997} and bulk metallic glasses \cite{lu2003} continues to
present great theoretical challenges. While dislocations have long
been recognized as playing a central role in plasticity of crystalline
systems, no counterpart is easily identifiable in disordered
matter. In addition, yield and flow \cite{liu2001} occur very far from
equilibrium, where the state of the system may have a complex history
dependence.

Progress towards understanding yield in glassy systems is currently
being made through a combination of simple models and particle-based
simulations. Falk and Langer's rate-equation-based Shear
Transformation Zone theory \cite{falk1998} was inspired in part by
observations made in molecular dynamics simulations.  Extensions of
this model \cite{lemaitre2001,falk2004} have been able to reproduce
many aspects of experiments on amorphous systems. A valuable
alternative approach is based on the energy landscape picture of
glasses, which relates well to the zero temperature, zero strain rate
limit of plasticity \cite{lacks2001,lemaitre2004}. Yet another
intriguing approach uses discretizations of continuum elasticity
theory to describe the long-range interactions of shear yielding
regions and resulting localization phenomena (shear
bands)\cite{picard2005}. A truly ``ab-initio'', but very challenging
approach to strained glasses is presently being pursued by extending
the mode-coupling theory of the glass transition to the effects of an
external drive \cite{fuchs2002, miyazaki2004, schweizer2005}.

Simulations of simple models of generic molecular glasses have played
an important role in developing the above theories of shear.  In
earlier work, we have shown that a generalized von Mises shear yield
criterion describes yield under general loading conditions as long as
failure is homogeneous \cite{rottler2001}.  We then examined the
dependence of the yield stress on rate and temperature
\cite{rottler2003}. The results were inconsistent with the simple, but
widely used Eyring model of viscoplasticity, which could not describe
the entire temperature range below the glass transition
temperature. These results suggest that in shearing glasses,
quantities other than the thermodynamic temperature contribute to the
activation of plastic events, although the precise nature of this
``effective temperature'' \cite {berthier2000} remains uncertain.

In addition to the above control parameters, experiments show that the
yield stress is sensitive to the age of the glass \cite{struik1978}.
Since glasses are out-of-equilibrium structures, they exhibit a slow,
but never ceasing, intrinsic aging dynamics
\cite{monthus1996,kob1997}.  In a simple picture, aging can be thought
of as thermally activated hopping in the glassy energy landscape
\cite{monthus1996}. With longer aging time, the system is able to pack
more densely and reach deeper and deeper energy minima. As a result,
the stress required to bring the system out of the local minimum into
a flowing configuration increases with waiting time
\cite{utz2000}. This scenario is reflected in the Soft Glassy Rheology
model, which predicts a slow logarithmic increase of the yield stress
with increasing aging time \cite{fielding2000}.

Our previous studies of shear yielding used a molecular glass that was
prepared through a rapid quench from the liquid state. The aging or
waiting time $t_w$ in the glassy regime before the application of
stress was typically shorter than the time to reach the yield point, a
situation not particularly common in experiments. The computational
effort to reach substantially longer waiting times used to be
prohibitive, but improvements in computing power are now making it
possible to study the relationship between aging and shear yielding in
molecular glasses. A first study of this kind was presented by Varnik
{\it et al.} \cite{varnik2004}, who showed for one fixed glassy
temperature and one fixed value of the strain rate that the yield
stress increases with age.  In the present work, we undertake a more
systematic study of shear yielding in the rate-age-temperature
parameter space and develop a phenomenological model that describes
the complex effects of all these parameters on the shear strength of
the glassy solid.

The methodology in the present work builds upon that of previous
studies \cite{rottler2001,rottler2003}. For consistency, we use the
80/20 binary Lennard Jones (LJ) mixture \cite{kob1995} that has been
employed extensively in molecular dynamics studies of glasses.  Our
units are the binding energy $u_0$ and length $a$ of the LJ potential
between majority particles.  The characteristic time is
$t_{LJ}=(ma^2/u_0)^{1/2}$, where $m$ is the mass of the particles.
When all interactions are truncated at particle separations greater
than $r_c=1.5 a$, the model exhibits a glass transition at a
temperature $T_g\approx 0.3 u_0/k_B$ \cite{rottler2003}.

\begin{figure}[t]
\begin{center}
\includegraphics[width=7cm]{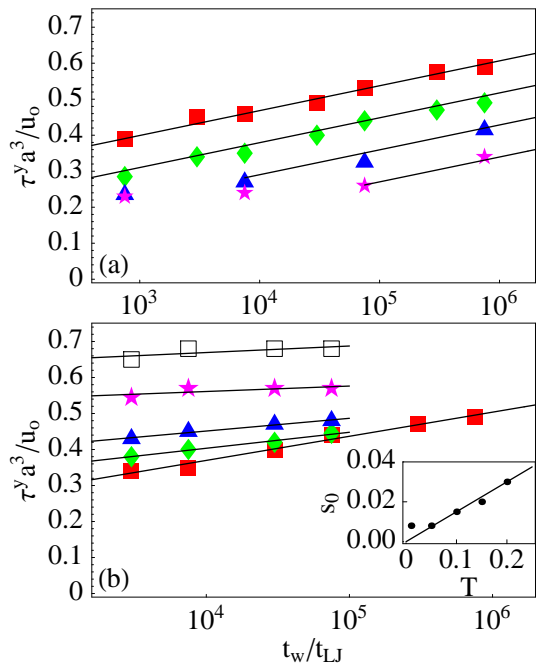}
\caption{\label{tauyvsage-fig} (Color online) Shear yield stress as a
function of waiting time.
(a) Four different rates
$\dot{\epsilon}=10^{-3} t_{LJ}^{-1}$ ($\blacksquare$),
$10^{-4} t_{LJ}^{-1}$ ($\blacklozenge$),
$10^{-5} t_{LJ}^{-1}$ ($\blacktriangle$)
and $10^{-6}t_{LJ}^{-1}$ ($\star$) at $T=0.2u_0/k_B$.
Solid line fits to the large $t_w$ region have common slope $s_0$
and successive lines are
separated by the vertical shift $s_1 \ln(10)$ predicted
by Eq. (\ref{eq:nobel}).
(b) Five different temperatures
$T=0.2u_0/k_B$($\blacksquare$),
$0.15 u_0/k_B$($\blacklozenge$),
$0.1 u_0/k_B$($\blacktriangle$),
$ 0.05 u_0/k_B$($\star$)
and $0.01 u_0/k_B$ ($\square$)
at fixed rate $ \dot{\epsilon}=10^{-4} t_{LJ}^{-1}$.
The solid lines are logarithmic fits to the data at
each $T$ and their slopes $s_0$ are plotted against
$T$ in the inset.
The straight line in the inset is a linear fit through the
origin.
}
\vspace{-1cm}
\end{center}
\end{figure}

We begin by preparing a melt configuration composed of 32768 particles
at an elevated temperature $T=1.3 u_0/k_B$ and then quench rapidly at
constant volume to a glassy temperature $T=0.2 u_0/k_B$ over a time of
$750 t_{LJ}$. The initial density is chosen so that the hydrostatic
pressure $p$ is close to zero at this temperature.  We then quench to
the desired temperature and wait a time $t_w$, while maintaining zero
pressure.  The density increases with waiting time in a logarithmic
fashion, in agreement with the intuitive idea that aging allows the
material to optimize its local packing \cite{struik1978}.  A 6\%
change is observed from $t_w=750$ to 750,000$t_{LJ}$ at $T=0.2
u_0/k_B$.  As expected for an unstrained system, the rate of
relaxation decreases with decreasing thermodynamic temperature,
becoming too small to detect at $T=0.01 u_0/k_B$.

After aging, a volume conserving shear is applied to the initially
cubic simulation cell. The strain along the z direction,
$\epsilon_{zz}$, increases at a constant rate $\dot{\epsilon}$, and
the strains along the two perpendicular directions are decreased
symmetrically to maintain fixed volume.  As in previous work
\cite{rottler2001}, we identify the shear yield stress $\tau^y$ with
the maximum of the deviatoric stress $\tau \equiv
[(\sigma_1-\sigma_2)^2+(\sigma_2-\sigma_3)^2+(\sigma_3-\sigma_1)^2]^{1/2}/3$,
where the $\sigma_i$ are the principal stresses.  The strain at yield,
$\epsilon^y$, is typically between 5 and 10\% for all cases studied.

Figure \ref{tauyvsage-fig}(a) shows $\tau^y$ as a function of waiting
time for four different strain rates from $\dot{\epsilon}=10^{-6}$ to
$10^{-3} t_{LJ}^{-1}$ at $T=0.2u_0/k_B$.  For all rates, the yield
stress increases logarithmically at long waiting times and the slope,
$s_0$, is independent of rate.  Note that the changes in $\tau^y$ are
too large to be explained by the increase in density with age
discussed above.  Thus the local internal structure in the glass must
also evolve logarithmically in time.  At short times and low shear
rates our results deviate from the logarithm, becoming nearly
independent of $t_w$.  This crossover is explained below
(Eq. \ref{eq:nobel}).

As shown in Figure \ref{tauyvsage-fig}(b), a logarithmic dependence on
waiting time is also observed at lower temperatures.  Data for a
relatively high rate, $10^{-4} t_{LJ}^{-1}$, is shown in order to
avoid the plateau seen at low $t_w$ and $\dot{\epsilon}$ in
Fig. \ref{tauyvsage-fig}(a).  The simplest picture of thermal
activation in an energy landscape would suggest that the slope, $s_0$,
scales linearly with temperature.  The inset to
Fig. \ref{tauyvsage-fig}(b) shows that our data are generally
consistent with $s_0 \propto T$.  The value for $T=0.01 u_0/k_B$ lies
above the linear fit, but the change in $\tau^y$ is very small at this
temperature and the data may be dominated by an initial transient.

\begin{figure}[t]
\begin{center}
  \includegraphics[width=7.5cm]{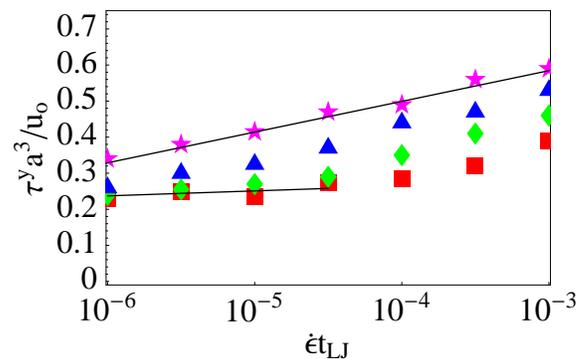}
\caption{\label{tauyvsrate-fig}(Color online) Shear yield stress as a
function of rate for 4 different waiting times $t_w=750 t_{LJ}$
($\blacksquare$), $7500 t_{LJ}$ ($\blacklozenge$),
$75000 t_{LJ}$ ($\blacktriangle$), and $750000 t_{LJ}$
($\star$) at $T=0.2 u_0/k_B$. Solid lines indicate logarithmic
fits with slopes of $s_1=0.037$ and $s'=0.006$ for long and short
waiting times, respectively.
Error bars are comparable to the symbol size.
}
\vspace{-1cm}
\end{center}
\end{figure}

We now turn to the combined effect of strain rate and aging on the
yield stress.  Figure \ref{tauyvsrate-fig} examines $\tau^y$ as a
function of strain rate over three orders of magnitude in waiting time
at $T=0.2 u_0/k_B$.  The data obtained with the shortest time,
$t_w=750 t_{LJ}$ (lowest curve), increase slowly with rate at low
rates and more rapidly at higher rates. This waiting time corresponds
to the one used in our earlier study of rate dependence
\cite{rottler2003}.  There, we showed that the small rate part of the
curve could be fitted to a logarithmic rate dependence with very
little change in the prefactor, $s'$, of the logarithm with $T$.  Here
we see that the region of rapid rate dependence moves to lower rates
as $t_w$ increases.  For the longest $t_w$, the entire curve can be
fit by a logarithmic rate dependence with a higher slope $s_1=0.037$.
We will refer to the regimes of low and steep slope as regimes I and
II below.  Note that there may also be a third regime at still higher
rates where the strain is faster than local elastic relaxations of the
solid.  As discussed in Ref.~\cite{rottler2003}, this may be reached
when $\dot{\epsilon}c/L\sim 1$, where $c$ is the speed of sound in the
glass and $L$ the scale of elastic heterogeneity.  However, this
regime is unlikely to be accessible to experiments.

Varnik {\it et al.} also observed a crossover between regimes I and II
that moved to lower rates with increasing waiting time
\cite{varnik2004}.  They argued that the crossover was associated with
shearing the system faster than its structural relaxation time, noting
evidence \cite{kob1997} that the time for atoms to escape from local
cages is comparable to $t_w$.  However, a data collapse motivated by
this picture did not describe their data.  Here we show that a
different physical picture, based only on the total effective age of
the system, can resolve these discrepancies and explain a wide range
of numerical data.

A logarithmic dependence on both waiting time and rate is commonly
observed in friction experiments \cite{dieterich1979,ruina1983}.  In
this context, phenomenological ``rate-state'' models have captured
many experimental results.  Such models assume that the response of
the system depends on both the rate of sliding and on a single ``state
variable,'' $\theta$, that increases with waiting time.  Replacing the
friction force and sliding velocity in these models by the yield
stress and strain rate yields:
\begin{equation}
\tau^y = \tau_0 + s_0 \ln (\theta) +
s_1 \ln (\dot{\epsilon}t_{LJ}) \ \ ,
\label{eq:ratestate}
\end{equation}
where the first logarithm reflects the growth in yield stress with
increasing age, and the second the increase in $\tau^y$ with shear
rate for a fixed state of the system.  An evolution equation for
$\theta$ must be specified to complete the model.  In this paper we
focus on the peak stress at the onset of yield, and it is reasonable
to write $\dot{\theta}=f(\epsilon_{zz},T)$.  One can choose the
normalization so $f(0,T)=1/t_{LJ}$.  Since the strain $\epsilon^y$ at
the onset of yield is nearly independent of shear rate and waiting
time, we can integrate the equation for $\theta$ to yield:
\begin{equation}
\tau^y = \tau_0 + s_0 \ln (t_w/t_{LJ} + \alpha/\dot{\epsilon}t_{LJ}) +
s_1 \ln (\dot{\epsilon}t_{LJ}) \ \ ,
\label{eq:nobel}
\end{equation}
where $\tau_0$, $s_0$ and $s_1$ are constants and $\alpha \equiv
\int_0^{\epsilon^y}d\epsilon_{zz} f(\epsilon_{zz},T)/f(0,T)$.  If $f$
is assumed to be independent of strain, $\alpha=\epsilon^y$, and
$\alpha/\dot{\epsilon}$ just corresponds to the time $t^y$ to strain
the system to yield.  If rejuvenation begins before $\epsilon^y$,
$\alpha$ will be smaller.  Strain may also accelerate aging
\cite{lacks2004} by lowering energy barriers, leading to larger values
of $\alpha$.

Equation (\ref{eq:nobel}) captures all of the limiting behavior seen
in our numerical results.  For fixed shear rate, only the first
logarithm is relevant.  At long waiting times, $\tau^y$ increases as
$s_0 \ln(t_w)$ with a rate independent slope $s_0$ and an offset that
rises as $s_1 \ln(\dot{\epsilon}t_{LJ})$.  This behavior is observed
in Fig. \ref{tauyvsage-fig}(a) at long $t_w$.  The saturation of $t^y$
at small $t_w$ arises because for $t_w < \alpha/\dot{\epsilon}$ the
state of the system is dominated by aging during the straining
interval.  For fixed waiting time, Eq.~(\ref{eq:nobel}) contains the
two regimes observed in Fig. \ref{tauyvsrate-fig}.  When $t_w$ is
smaller than $\alpha/\dot{\epsilon}$ (regime I), $t_w$ is irrelevant
and the two logarithms compete.  A higher rate increases the intrinsic
strength through the second logarithm in Eq.~(\ref{eq:nobel}), but
allows less time for aging to increase the yield stress through the
first logarithm.  The net result is that $\tau^y$ rises as $s'
\ln(\dot{\epsilon})$, where $s'=s_1-s_0$ is much smaller than either
$s_1$ or $s_0$.  For large $t_w$ (regime II), only the second
logarithm in Eq.~(\ref{eq:nobel}) contributes and $\tau^y$ rises as $s_1
\ln(\dot{\epsilon})$.  Here, the solid is strained so rapidly that it
does not age significantly before yield occurs.

\begin{figure}[t]
\begin{center}
\includegraphics[width=7.5cm]{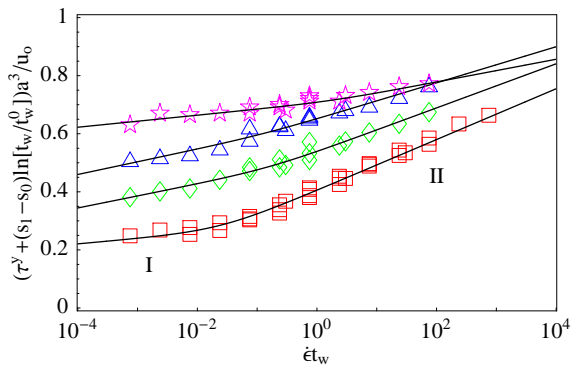}
\caption{\label{collapse-fig}(Color online) Plot of data for all
$t_w$ and $\dot{\epsilon}$ at
$T=0.2 u_0/k_B$ ($\square$),
$0.1 u_0/k_B$ ($\lozenge$),
$0.05 u_0/k_B$ ($\triangle$)
and $0.01 u_0/k_B$ ($\star$),
and universal curves (solid lines) predicted by
Eq. (\ref{eq:nobel}).
}
\end{center}
\vspace{-0.5cm}
\end{figure}

Equation (\ref{eq:nobel}) also implies that data for all waiting times
and shear rates should collapse onto a universal curve if $\tau^y + s'
\ln(t_w/t_w^0)$ is plotted against $\dot{\epsilon} t_w$, where $t_w^0$
is any reference time.  Figure~\ref{collapse-fig} shows the success of
this collapse over the whole temperature range.  There are no
adjustable parameters in the collapse, since $s'=s_1-s_0$ was
determined from separate measurements of $s_1$ and $s_0$ in the
asymptotic regimes of plots like Figs.~\ref{tauyvsage-fig} and
\ref{tauyvsrate-fig}.

The solid lines in Fig.~\ref{collapse-fig} show the predictions of
Eq.~(\ref{eq:nobel}).  These lines do require fits to $\tau_0$ and
$\alpha$ for each $T$.  The crossover between regimes I and II occurs
when $\dot{\epsilon} t_w \approx \alpha$, and $\alpha$ clearly
increases by more than an order of magnitude with decreasing
temperature.  For $T=0.2u_0/k_B$, the best fit gave $\alpha=0.02$ with
an uncertainty of about a factor of two due to the errorbars on $s_0$
and $s_1$.  Note that this range of $\alpha$ is comparable to
$\epsilon^y$, implying that the rate of aging during strain is
comparable to that at zero strain.  The increase in $\alpha$ with
decreasing $T$ implies that the aging is accelerated by strain at low
temperatures.  Our previous studies of the rate of structural
rearrangements in strained and unstrained glasses \cite{rottler2003}
support these conclusions.  The probability distribution of sudden
local stress and strain changes was monitored as a function of the
applied strain at different $T$.  Very near $T_g$, the rate and
magnitude of local changes is nearly independent of the applied
strain, explaining why $\alpha \approx \epsilon^y$.  At low $T$, the
probability of large events increases rapidly with strain, becoming
comparable to the rate at $T_g$ near the yield point.  In this regime,
most of the thermal activation occurs at strains near the yield point
leading to $\alpha \gg \epsilon^y$.

Equation (\ref{eq:nobel}) can also be obtained from a simple
modification of Eyring's model \cite{eyring1936} of stress-assisted
thermal activation over energy barriers of height $\Delta E$.  In this
very simple but commonly used model, the strain rate is associated
with transition rates over barriers whose height decreases linearly
with applied stress.  One obtains $\tau^y= \Delta E/V^* +
(k_BT/V^*)\ln\left[\dot{\epsilon}/\nu_0\right]$, where $V^\star$ is a
constant called the ``activation volume'' and $\nu_0$ an attempt
frequency. This model provides a basic explanation for logarithmic
rate behavior, but does not include aging.  A suitable, albeit ad hoc,
extension is to include an increase in $\Delta E$ with the total age
of the system, as for instance $\Delta E=\Delta E_0 + f(T) \ln{[t_w +
\alpha/\dot{\epsilon}]}$.  This immediately yields
Eq.~(\ref{eq:nobel}) with $\tau_0={\Delta E}/V^*$, $s_0 = f(T)/V^*$
and $s_1 = k_B T/V^*$.  However, the observed temperature dependence
of $s_0$ and $s_1$ does not follow simply from these relations.  While
$s_0$ is approximately proportional to temperature
(Fig.~\ref{tauyvsage-fig}(b) inset), $s_1$ varies slowly at high $T$
and appears to approach a constant at low $T$.  This would require
$V^* \rightarrow 0$ as $T \rightarrow 0$.  It seems more likely that
$s_1$ is related to intrinsic rate effects.  Several analytic models
\cite{lemaitre2001,falk2004,fuchs2002,schweizer2005,fielding2000}
include such effects, but their consequences have not been worked out
for the entire rate-temperature-age parameter space.

The data described above are all for the onset of yield from an
unstrained state.  It is interesting to compare them to previous
studies of the flow stress in steady-state shear
\cite{rottler2003,varnik2004}.  Sheared systems can not reach regime
II of Fig. ~\ref{collapse-fig}, because they are constantly being
``rejuvenated'' \cite{utz2000,liu2001}.  One expects that the
effective waiting time should scale with the inverse shear rate,
leading to logarithmic rate dependence with slope $s_1-s_0$
characteristic of regime I.  The measured slope is indeed closer to
that of regime I and is also relatively insensitive to temperature
\cite{rottler2003,varnik2004}.

In conclusion, we have found a complex interplay of waiting time,
temperature and rate in determining the yield stress of glassy solids.
In the absence of an imposed strain, the system only evolves through
thermal activation.  Aging leads to a logarithmic increase in density
with a prefactor that decreases rapidly as $T$ decreases.  The state
of the system continues to evolve through thermal activation during
shear.  The yield stress reflects both this evolution and intrinsic
rate dependence.  A unified description (Eq.~(\ref{eq:nobel})) based
on rate-state models of friction \cite{dieterich1979,ruina1983} is
able to collapse all data at each temperature onto a universal curve
(Fig.~\ref{collapse-fig}).  At large values of $\dot{\epsilon} t_w$
(regime II), there is little evolution of the system as it is strained
to yield and $\tau^y$ rises rapidly with strain rate.  At small values
of $\dot{\epsilon} t_w$ (regime I), the stress rises less rapidly with
$\dot{\epsilon}$ because the increase in stress with rate is partially
offset by a reduction in the time for aging.  The model can be
obtained from a simple modification of the Eyring model, and the
results may help test and motivate future analytic theories of
plasticity in glassy materials.

%\bibliography{glasses} 

\end{document}